\documentclass[3p,times,twocolumn]{elsarticle}

\usepackage{ecrc}

\volume{00}

\firstpage{1}

\journalname{Nuclear Physics B Proceedings Supplement}

\runauth{}

\jid{nuphbp}

\jnltitlelogo{Nuclear Physics B Proceedings Supplement}

\usepackage{amssymb}

\usepackage[figuresright]{rotating}

\begin{document}

\begin{frontmatter}



\dochead{}

\title{The LHAASO experiment: from Gamma-Ray Astronomy to Cosmic Rays}


\author{G. Di Sciascio on behalf of the LHAASO Collaboration}

\address{INFN, Sezione Roma Tor Vergata, Roma, Italy}

\begin{abstract}
 LHAASO is expected to be the most sensitive project to face the open problems in Galactic cosmic ray physics through a combined study of photon- and charged particle-induced extensive air showers in the energy range 10$^{11}$ - 10$^{17}$ eV.
This new generation multi-component experiment will be able of continuously surveying the gamma-ray sky for steady and transient sources from about 100 GeV to PeV energies, thus opening for the first time the 10$^2$--10$^3$ TeV range to the direct observations of the high energy cosmic ray sources.

In addition, the different observables (electronic, muonic and Cherenkov components) that will be measured in LHAASO will allow the study of the origin, acceleration and propagation of the radiation through a measurement of energy spectrum, elemental composition and anisotropy with unprecedented resolution. 
The installation of the experiment started at very high altitude in China (Daocheng site, Sichuan province, 4410 m a.s.l.). 
The commissioning of one fourth of the detector will be implemented in 2018.
The completion of the installation is expected by the end of 2021.
\end{abstract}

\begin{keyword}
Cosmic Rays Physics \sep Gamma Ray Astronomy \sep LHAASO experiment 


\end{keyword}

\end{frontmatter}



\section{Introduction}
The Large High Altitude Air Shower Observatory (LHAASO) project is a new generation instrument, to be built at 4410 meters of altitude in the Sichuan province of China, with the aim of studying with unprecedented sensitivity the energy spectrum, the elemental composition and the anisotropy of cosmic rays (hereafter CR) in the energy range between 10$^{12}$ and 10$^{17}$~eV, as well as to act simultaneously as a wide aperture ($\sim$2 sr), continuosly-operated gamma ray telescope in the energy range between 10$^{11}$ and $10^{15}$~eV.
The remarkable sensitivity of LHAASO in CR physics and gamma astronomy would play a key-role in the comprehensive general program to explore the \emph{``High Energy Universe''}.

The first phase of LHAASO will consist of the following major components:
\begin{itemize}
\vspace{-0.2cm}
\item 1 km$^2$ array (LHAASO-KM2A) for electromagnetic particle detectors (ED) divided into two parts: a central part including 4931 scintillator detectors 1 m$^2$ each in size (15 m spacing) to cover a circular area with a radius of 575 m and an outer guard-ring instrumented with 311 EDs (30 m spacing) up to a radius of 635 m.
\vspace{-0.2cm}
\item An overlapping 1 km$^2$ array of 1146 underground water Cherenkov tanks 36 m$^2$ each in size, with 30 m spacing, for muon detection (MD, total sensitive area $\sim$42,000 m$^2$).
\vspace{-0.2cm}
\item A close-packed, surface water Cherenkov detector facility with a total area of about 78,000 m$^2$ (LHAASO-WCDA).
\vspace{-0.2cm}
\item 12 wide field-of-view air Cherenkov telescopes (LHAASO-WFCTA).
\end{itemize}

LHAASO will be located at high altitude (4410 m asl, 600 g/cm$^2$, 29$^{\circ}$ 21' 31'' N, 100$^{\circ}$ 08'15'' E) in the Daochen site, Sichuan province, P.R. China. 
The commissioning of one fourth of the detector will be implemented in 2018.
The completion of the installation is expected by the end of 2021.\\

LHAASO will enable studies in CR physics and gamma-ray astronomy that are unattainable with the current suite of instruments:
\begin{itemize}
\item[1)] LHAASO will perform an \emph{unbiased sky survey of the Northern sky} with a detection threshold of a few percent Crab units at sub-TeV/TeV energies and around 100 TeV in one year. This sensitivity grants a high discovery potential of flat spectrum Geminga-like sources not observed at GeV energies.
This unique detector will be capable of continuously surveying the $\gamma$-ray sky for steady and transient sources from about 100 GeV to 1 PeV. \\
From its location LHAASO will observe at TeV energies and with high sensitivity about 30 of the sources catalogued by Fermi-LAT at lower energy, monitoring the variability of 15 AGNs (mainly blazars) at least.
\vspace{-0.2cm}
\item[2)] The sub-TeV/TeV LHAASO sensitivity will allow to observe AGN flares that are unobservable by other instruments, including the so-called TeV orphan flares. 
\vspace{-0.2cm}
\item[3)] LHAASO will study in detail the high energy tail of the spectra of most of the $\gamma$-ray sources observed at TeV energies, opening for the first time the 100--1000 TeV range to the direct observations of the high energy cosmic ray sources.
\emph{When new wavelength bands are explored in astronomy, previously unknown sources and unknown types of sources are discovered. LHAASO's wide field of view provides therefore a unique discovery potential.}
\vspace{-0.2cm}
\item[4)] LHAASO will map the Galactic \emph{diffuse gamma-ray emission} above few hundreds GeV and thereby measure the CR flux and spectrum throughout the Galaxy with high sensitivity. 
The measurement of the space distribution of diffuse $\gamma$-rays will allow to trace the location of the CR sources and the distribution of interstellar gas.
\vspace{-0.2cm}
\item[5)] The high background rejection capability in the 10 -- 100 TeV range will allow LHAASO to measure the \emph{isotropic diffuse flux of ultrahigh energy $\gamma$ radiation} expected from a variety of sources including Dark Matter and the interaction  of 10$^{20}$ eV CRs with the 2.7 K microwave background radiation. 
In addition, LHAASO will be able to achieve a limit below the level of the IceCube diffuse neutrino flux at 10 -- 100 TeV, thus constraining the origin of the IceCube astrophysical neutrinos.
\vspace{-0.2cm}
\item[6)] LHAASO will allow the reconstruction of the energy spectra of different CR mass groups in the 10$^{12}$ -- 10$^{17}$ eV with unprecedented statistics and resolution, thus tracing the light and heavy components through the knee of the all-particle spectrum.
\vspace{-0.5cm}
\item[7)] LHAASO will allow the measurement, for the first time, of the CR anisotropy across the knee separately for light and heavy primary masses.
\vspace{-0.2cm}
\item[8)] The different observables (electronic, muonic and Cherenkov components) that will be measured in LHAASO will allow a detailed investigation of the role of the hadronic interaction models, therefore investigating if the EAS development is correctly described by the current simulation codes.
\vspace{-0.2cm}
\item[9)] LHAASO will look for signatures of WIMPs as candidate particles for DM with high sensitivity for particles masses above 10 TeV. Moreover, axion-like particle searches are planned, where conversion of gamma-rays to/from axion-like particles can create distinctive features in the spectra of gamma-ray sources and/or increase transparency of the universe by reducing the Extragalactic Background Light (EBL) absorption. 
Testing of Lorentz invariance violation as well as the search for Primordial Black Holes and Q--balls will also be part of the scientific programme of the experiment.
\end{itemize} 

In the next decade CTA-North and LHAASO are expected to be the most sensitive instruments to study Gamma-Ray Astronomy in the Northern hemisphere from about 20 GeV up to PeV.

\section{Scientific Motivation}

\subsection{Open problems in Galactic Cosmic Ray Physics}
\label{cr}

The riddle of the origin of CR is open since more than one century. The identification of the galactic sources able to accelerate particles beyond PeV (=10$^{15}$ eV) energies is certainly one of the main open problems. 
Energy spectrum, elemental composition and anisotropy in the CR arrival direction distribution are three basic observables  crucial for understanding origin, acceleration and propagation of the radiation.
\begin{table*}[h]
\label{tab:one}
{
\footnotesize 
\centerline{\bf Table 1: Characteristics of different EAS-arrays}
\vspace{0.2cm}
\begin{center}
{\begin{tabular}{|c|c|c|c|c|}
\hline
  Experiment &   Altitude (m) & e.m. Sensitive Area & Instrumented Area & Coverage \\
 & & (m$^2$) & (m$^2$) & \\
\hline
 LHAASO & 4410 & 5.2$\times$10$^3$ & 1.3$\times$10$^6$ & 4$\times$10$^{-3}$ \\
 \hline
 TIBET AS$\gamma$ & 4300 & 380 & 3.7$\times$10$^4$ & 10$^{-2}$ \\
 \hline
IceTop & 2835 & 4.2$\times$10$^2$ & 10$^6$ & 4$\times$10$^{-4}$ \\
\hline
ARGO-YBJ & 4300 & 6700 & 11,000 & 0.93 (central carpet)\\
\hline
KASCADE & 110 & 5$\times$10$^2$ & 4$\times$10$^4$ & 1.2$\times$10$^{-2}$ \\
\hline
KASCADE-Grande & 110 & 370 & 5$\times$10$^5$ & 7$\times$10$^{-4}$ \\
\hline
CASA-MIA & 1450 & 1.6$\times$10$^3$ & 2.3$\times$10$^5$ & 7$\times$10$^{-3}$ \\
\hline
\hline
 & & $\mu$ Sensitive Area & Instrumented Area & Coverage \\
  & & (m$^2$) & (m$^2$) & \\
\hline
LHAASO & 4410 & 4.2$\times$10$^4$ & 10$^6$ & 4.4$\times$10$^{-2}$ \\
 \hline
 TIBET AS$\gamma$ & 4300 & 4.5$\times$10$^3$ & 3.7$\times$10$^4$ & 1.2$\times$10$^{-1}$\\
 \hline
 KASCADE & 110 & 6$\times$10$^2$ & 4$\times$10$^4$ & 1.5$\times$10$^{-2}$ \\
 \hline
 CASA-MIA & 1450 & 2.5$\times$10$^3$ & 2.3$\times$10$^5$ & 1.1$\times$10$^{-2}$ \\
\hline
\end{tabular} }
\end{center}
}
\end{table*}

In the PeV energy region an accurate measurement of the CR primary spectrum can be carried out only by ground-based Extensive Air Showers (hereafter EAS) arrays.
In fact, since the CR flux rapidly decreases with increasing energy and the size of detectors is constrained by the weight that can be placed on satellites/balloons, their collecting area is small and determine a maximum energy (of the order of 100 TeV/nucleon) related to a statistically significant detection. In addition, the limited volume of the detectors make difficult the containement of showers induced by high energy nuclei, thus limiting the energy resolution of instruments in direct measurements.

Understanding the CR origin and propagation at any energy is made difficult by the poor knowledge of the elemental composition of the radiation as a function of the energy.
Last generation experiments, measuring with high resolution different EAS components (mainly the number of electrons, N$_e$, and the number of muons, N$_\mu$, at observation level), have reached the sensibility to separate two mass groups (light and heavy) with an analysis technique not critically based on EAS simulations or five mass groups (H, He, CNO, MgSi, Fe) with an unfolding technique that is heavily based on simulations. 

In the standard picture, mainly based on the results of the KASCADE esperiment, the knee is attributed to the steepening of the p and He spectra \cite{kascade}. 
However, a number of experiments (in particular those obtained by experiments located at high altitudes) seem to indicate that the bending of the light component (p+He) is well below the PeV and the knee of the all-particle spectrum is due to heavier nuclei \cite{tibet,casamia,basje-mas}. 
Recent results obtained by the ARGO-YBJ experiment (located at 4300 m asl) reported evidence, with different analyses, that the knee of the light component starts at $\sim$700 TeV, well below the knee of the all-particle spectrum that is confirmed by ARGO-YBJ at about 3$\cdot$10$^{15}$ eV \cite{hybrid15}.

A large number of theoretical papers discussed the highest energies achievable in SNRs and the possibility that protons can be accelerated up to PeVs \cite{blasi13}.
\emph{The determination of the proton knee, as well as the measurement of the evolution of the heavy component across the knee, are the key components for understanding CR acceleration mechanisms and the propagation processes in the Galaxy, and to investigate the transition from Galactic to extra-galactic CRs.}

At higher energies, KASCADE-Grande, IceTop and Tunka experiments observed a hardening slightly above 10$^{16}$ eV and a steepening at log10(E/eV) = 16.92$\pm$0.10 in the CR all-particle spectrum.
A steepening at log10(E/eV) = 16.92$\pm$0.04 in the spectrum of the electron poor event sample (heavy primaries) and a hardening at log10(E/eV) = 17.08$\pm$0.08 in the electron rich (light primaries) one were observed by KASCADE-Grande even if with modest statistical significance \cite{kascadeg-chiavassa}. The absolute fluxes of CRs with different masses measured by KASCADE-Grande are however strongly dependent on the adopted hadronic interaction models \cite{kascadeg-hadron}, thus requiring new high resolution data to clarify the observations.

The measurement of the anisotropy in the arrival direction distribution of CRs is a complementary way to understand the origin and propagation of the radiation. It is also a tool to probe the structure of the magnetic fields through which CRs travel.
From the theoretical viewpoint, the anisotropy of CRs is important as a direct trace of potential sources. 
Moreover, the study of the anisotropy can clarify the origin of the knee. Indeed, if the knee is due to an increasing inefficiency in CR containment in the Galaxy a change in anisotropy is expected. If, on the contrary, the knee is related to the limit of the acceleration mechanism, we do not expect a change of anisotropy across the knee.

Data reveal characteristics of the CR anisotropy that cannot be described by any standard diffusion model of CR propagation in the interstellar medium, showing that the propagation of CR inside the Galaxy is not well-understood yet.
The origin of the observed anisotropies is still unknown. So far, no theory of CRs in the Galaxy exists yet to explain the observations leaving the standard model of CRs and that of the local magnetic field unchanged at the same time \cite{disciascio13,disciascio15}.

A measurement of the evolution of anisotropy with the energy across the knee and the determination of the chemical nature of excesses and deficits, are crucial to disentangle between different models of CR propagation in the Galaxy.
Only high resolution and high statistics experiments will further increase our knowledge about energy spectrum, elemental composition and anisotropies of Galactic CRs.

The LHAASO experiment satisfies both requirements.
In this project a km$^2$ area will be instrumented with a high coverage factor for the detection of the electromagnetic and muonic EAS components. 
In addition, LHAASO will be instrumented also with an array of wide field of view (FoV) Cherenkov telescopes to measure the CR  energy spectrum with good energy resolution.

In the Table \ref{tab:one} the characteristics of the LHAASO KM2A array are compared with other experiments. As can be seen, LHAASO will operate with a coverage of $\sim$0.4\% over more than 1 km$^2$ area.
The sensitive area of muon detectors is unprecedented and is about 70 times larger than KASCADE, with a coverage of $\sim$4\% over 1 km$^2$.

One of the most important characteristics of LHAASO is the high altitude location.
Working at high altitude (4410 m asl) is fruitful because: (1) the shower fluctuations are reduced since the detector approaches 
the depth of the EAS maximum longitudinal development; (2) the average e.m. shower size in the knee energy region is almost the same for all nuclei, providing a composition independent estimator of the energy; (3) the low energy threshold allows the calibration of the absolute energy scale exploiting the \emph{"Moon shadow"} technique, as demonstrated by the ARGO-YBJ experiment  \cite{argo-moon}. In addition the overposition with direct measurements allows a cross-calibration up to about 100 TeV/nucleon. The calibration of the relation between shower size and primary energy is one of the most important problem for ground-based measurement, heavely affecting the reconstruction of the CR energy spectrum.

The key point of future experiments aiming at studying the cosmic radiation is the possibility to separate, on a event by event basis, as much as possible mass groups to measure their spectra and large scale anisotropies. 
As demonstrated in the hybrid measurement carried out with ARGO-YBJ, the array of Cherenkov telescopes will allow the selection, with high resolution, of the main primary mass groups on an event-by-event basis, without any unfolding procedure and the recostruction of energy spectra with an energy resolution of order of 20\% \cite{hybrid15}.
In addition, the correlation between electromagnetic, muonic and Cherenkov components will allow the study of the dependence upon different hadronic models thus investigating if the EAS development is correctly described by the current simulation codes.

\subsection{Motivation for a Multi-TeV ($>$10 TeV) $\gamma$-ray telescope}
\label{ga}

High-energy gamma-ray (and neutrino) observations are an essential probe of CRs, if gamma rays (and neutrinos) are produced by CRs interacting close to their sources. 
In fact photons (and neutrinos) travel in straight lines, unperturbed by the magnetic fields, and, unlike the charged CRs, point back to their sources, providing the direction to the CR accelerator. 
The integrated study of charged CRs and of gamma rays and neutrinos is one of the most important (and exciting) fields in the so-called \emph{'multi-messenger astronomy'}.

Presently there is a general consensus that CRs with energy up to the knee of the all-particle spectrum  ($\sim$3$\times$10$^{15}$ eV), and probably even up to 10$^{17}$ eV, are mainly Galactic, produced and accelerated by the shock waves of Supernova Remnants (SNR) expanding shell \cite{aha13}.
Recently AGILE and Fermi observed GeV photons from two young SNRs (W44 and IC443) showing the typical spectrum feature around 1 GeV (the so called \emph{'$\pi^0$ bump'}, due to the decay of $\pi^0$) related to hadronic interactions \cite{pizero-a,pizero-f}. 
This important measurement however does not demonstrate the capability of SNRs to accelerate CRs up to the knee of the all-particle spectrum and above. 
In fact, unlike neutrinos that are produced only in hadronic interactions, the question whether $\gamma$-rays are produced by the decay of $\pi^0$ from protons or nuclei interactions (\emph{'hadronic'} mechanism), or by a population of relativistic electrons via Inverse Compton scattering or bremsstrahlung (\emph{'leptonic'} mechanism), still need a conclusive answer.

Gamma-ray astronomy at and above 100 TeV is of extreme importance since
\begin{itemize}
\item[(a)] The Inverse Compton scattering at these energies is strongly suppressed by the Klein-Nishina effect. Therefore, the observation of a $\gamma$-ray power law spectrum with no break up to the 100 TeV range would be a sufficient condition to proof the hadronic nature of the interaction.
\item[(b)] PeV protons colliding with molecular clouds or other matter produce $\sim$100 TeV gamma rays. The survey of the sky at 100--1000 TeV photon energies is crucial to face the main open problem of high energy astrophysics: the identification of CRs sources, the so-called \emph{'PeVatrons'}.
\end{itemize}
In the last two decades a large number of gamma-ray sources (more than 3000 in the Fermi catalogue) have been observed by space borne and ground based telescopes. About 170 of them emit radiation up to TeV energies. 
These sources belong to different classes of objects, both galactic (pulsars, pulsar wind nebulae PWN, supernova remnants SNR, compact binary systems, etc.) and extragalactic (active galactic nuclei AGN and gamma ray bursts GRB), in which the emission of high energy photons can be produced by different mechanisms. 
About 1/3 of galactic sources are still unidentified and $\sim$70\% are extended.

So far, no photons of energy above 100 TeV have been observed from any source, and only six sources have data above 30 TeV: 
the SNR RX J1713.7-3946, the PWN Crab and Vela-X, and the extended sources MGROJ2031+41, MGROJ2019+37 and MGROJ1908+06, all of them probably PWN too.
Their spectra above 30 TeV is however known with large uncertainties, being the sensitivity of current instruments at the highest energies not enough to determine clearly the spectral shape. 

To open the 10$^2$--10$^3$ TeV range to observations a very large effective area is required.
The most sensitive experimental technique for the observation of multi-TeV $\gamma$-rays is the detection of EAS via large ground-based arrays.
The muon content of photon-induced showers is very low, therefore these events can be discriminated from the large background of CRs via a simultaneous detection of muons that originate in the muon-rich CR showers. 
Their large FoV and high duty cycle ($>$ 90\%) make these observational technique particularly suited to perform unbiased all-sky surveys (not simply of limited regions as the Galactic plane) and to monitor the sky for the brightest transient emission from AGN and GRB, and search for unknown transient phenomena. 
The highest energy gamma rays and the shortest timescales of variability provide the strongest constraints on the acceleration mechanisms at work in these sources.
\begin{figure}
  \begin{center}
 \includegraphics[width=0.8\columnwidth]{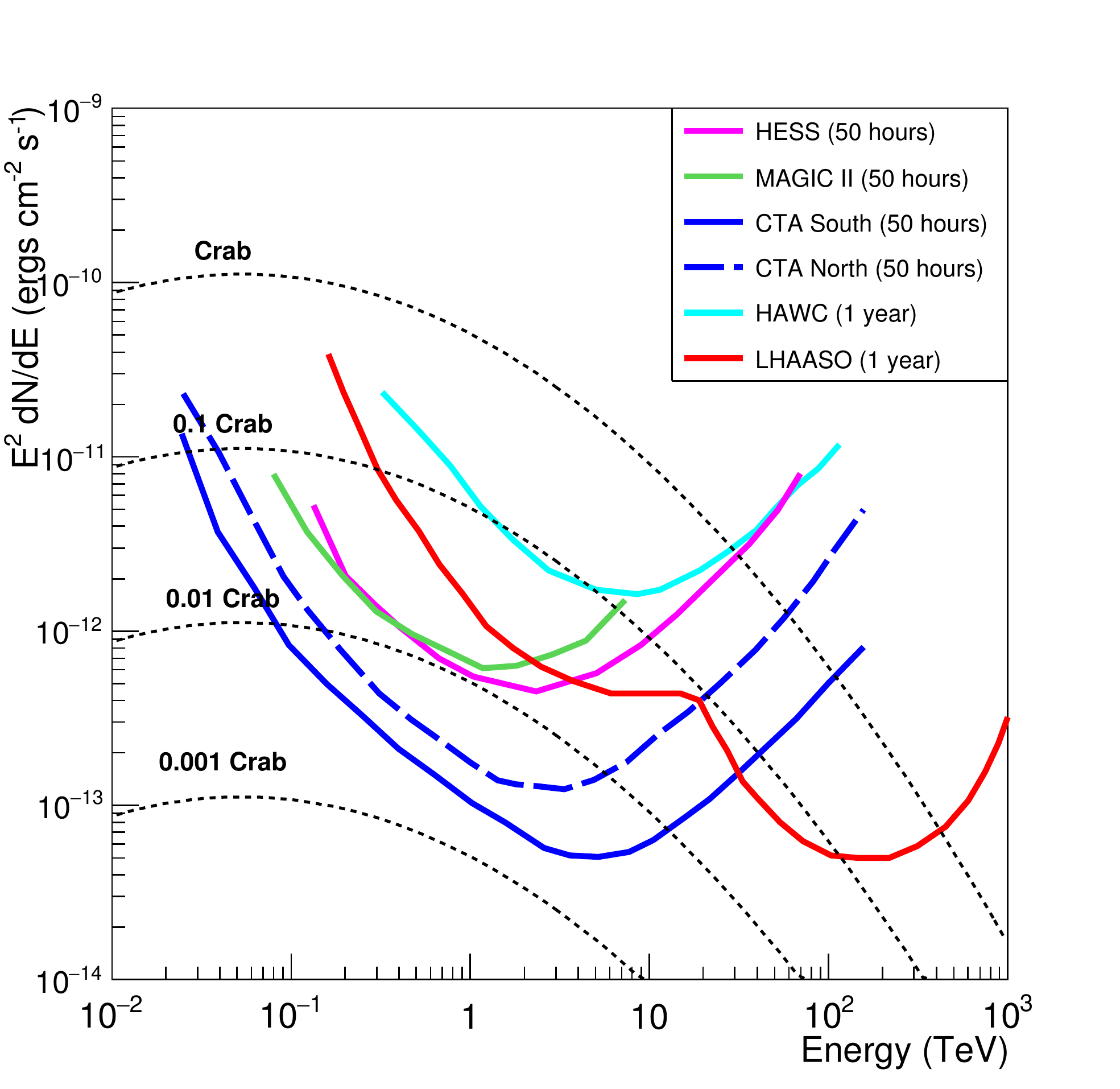}
    \caption{Differential sensitivity of LHAASO to a Crab-like point gamma ray sources compared to other experiments (multiplied by E$^2$). The Crab Nebula spectrum, extrapolated to 1 PeV, is reported as a reference together with the spectra corresponding to 10$\%$, 1$\%$ and 0.1$\%$  of the Crab flux.}
    \label{fig:lhaaso_sens}
  \end{center}
\end{figure}

It is known that the ensembles of the brightest sources in the GeV and TeV ranges do not coincide exactly, with some bright sources detected by Fermi relatively faint at TeV energy and vice-versa. It is therefore possible that the brightest sources in the 100 TeV ranges are unexpected. In this sense the large sky coverage of an EAS detector is well suited to discover the emission from unknown sources. LHAASO's wide field of view provides a unique discovery potential.

It has also been proposed, that the highest energy particle produced in astrophysical accelerators can escape rapidly the accelerator, and therefore that the highest energy emission is not from point-like or quasi point-like sources, but it could have a broader extension, mapping the interstellar gas distribution in the vicinity of the accelerators, and perhaps also the structure of the regular magnetic field in the region, if the diffusion is strongly anisotropic.
In these circumstances the identification of the PeVatrons is more difficult, but the wide FoV of the EAS-arrays could be decisive  to trace the $\gamma$-ray emission.

LHAASO, thanks to the large area of the km$^2$ array and the high capability of background rejection, can reach sensitivities above 30 TeV range $\sim$100 times higher than that of current instruments, offering the possibility to monitor for the first time the $\gamma$-ray sky up to PeV energies.
In addition, at sub-TeV and TeV energies LHAASO will continuously observe all the Northern flaring gamma-ray sky with a sensitivity of a few percent of the Crab Nebula flux.

The sensitivity of LHAASO to point--like gamma-ray sources is shown in Fig. \ref{fig:lhaaso_sens} where is compared to other experiments. The sensitivity curve has been calculated for a Crab-like energy spectrum extending to PeV without any cutoff. The LHAASO sensitivity curve shows a structure with two minima, reflecting the fact that the observation and identification of photon showers in different energy ranges is controlled by different detectors: the water Cherenkov array (WCDA) in the range $\sim$0.1 -- 10~TeV and KM2A array above 10 TeV \cite{cui2014}. 

For comparison of sensitivities, it is important to note that the LHAASO sensitivity shown is the point-source \emph{survey} sensitivity for $\sim$1.5 sr of the sky. 
While the sensitivities of EAS-arrays are also valid for surveys, the sensitivities of Cerenkov telescopes are given for pointed observations of 50 h 'on source' in a small FoV of the order of $\pi$/100 sr. 

The choice of different conventions is inevitable because of the different operation modes of the two detection techniques.
Cherenkov telescopes work only during clear moonless nights, with a total observation time of about 1000--1500 hours per year
(depending on the location), and have a FoV of a few degrees of radius. This implies that they can observe only one 
(or very few) sources at the same time, and only in the season of the year when the source culminates during night time.
Fifty hours is a typical time that a Cherenkov telescopes dedicate to a selected source in one year and $\sim$10 hours are as a rule devoted to survey observations.

In contrast, the sky region observed by an EAS detector is completely determined by its geographical location.
The detector observes nearly continuously a large fraction of the celestial sphere (spanning 360 degrees in right ascension and about 90 degrees in declination).
Sources located in this portion of the sky are in the FoV of the detector, either always, or for several hours per day, depending on their celestial declination.
This situation is ideal to perform sky surveys, discover transients or explosive events (such as GRBs), and monitor variable or flaring sources such as AGNs.

As can be seen in Fig. \ref{fig:lhaaso_sens}, at energies above few TeV the LHAASO one year sensitivity is better than the sensitivity of a 50 hours MAGIC or HESS observation of a single source.
An IACT can only spend up to $\sim$200 hours per year observing a single interesting source due to solar and lunar constraints. However, even with maximum exposure, an IACT still wouldn't be able to match LHAASO sensitivity above few tens TeV.
\begin{figure}
  \begin{center}
 \includegraphics[width=0.8\columnwidth]{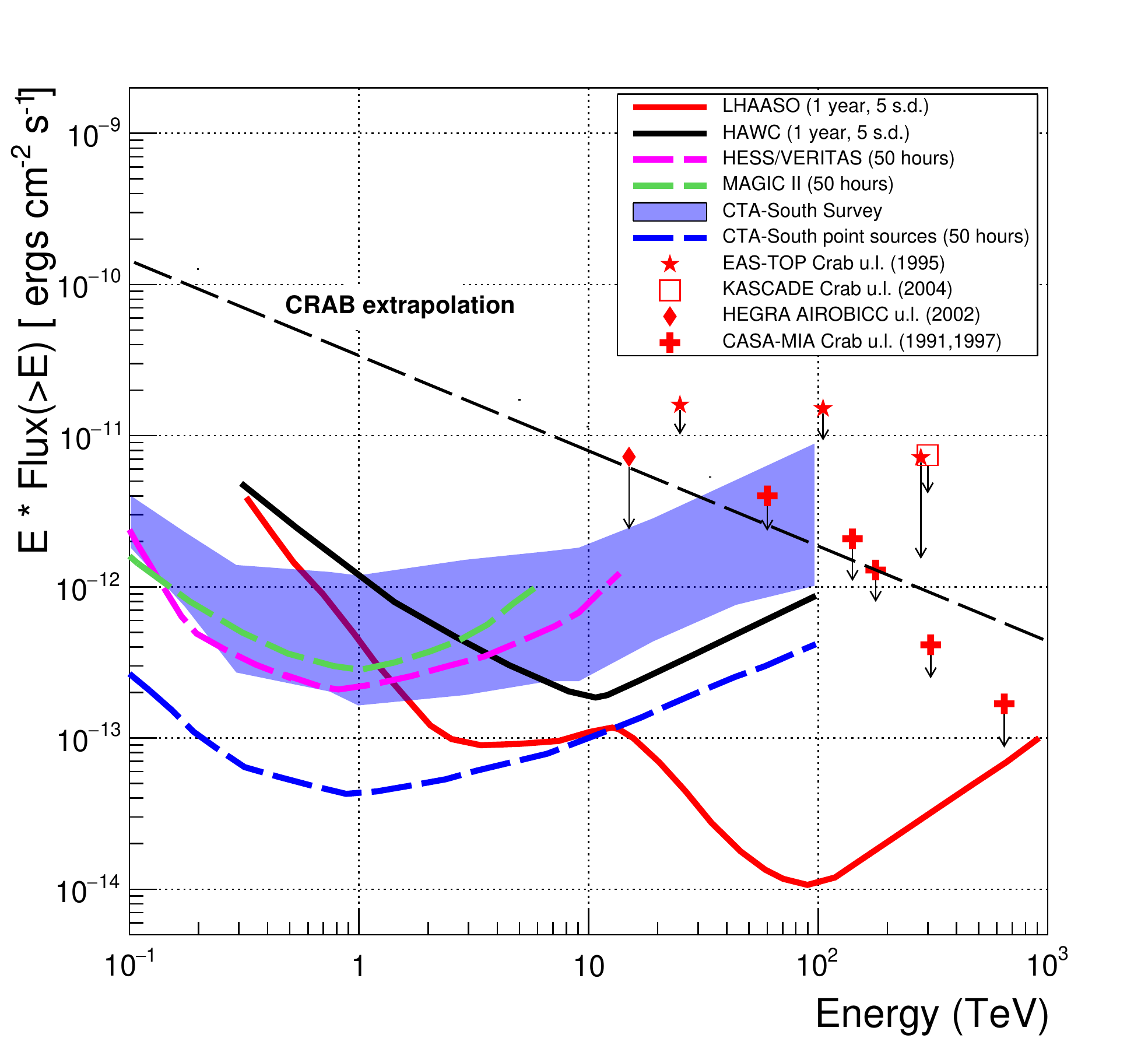}
\caption{Integral sensitivity of LHAASO as a function of the energy compared to HAWC, HESS, MAGIC II and CTA-South sensitivities. The CTA-South sensitivity to sky survey \cite{cta-survey,hiscore} is also shown.    
Upper limits to ultra high-energy gamma-ray emission set by different experiments in the Northern hemisphere are also reported \cite{eastop97,kascade04,hegra02,casa-mia97}.} 
    \label{fig:lhaaso-skysurvey}
  \end{center}
\end{figure}

One of the most interesting aspect of LHAASO is its large FoV and the capability to monitor every day large portion of the sky.
If we consider a survey of the Galactic Plane in the latitude band $-4^{\circ} < b < +4^{\circ}$, for a longitude interval of 200$^{\circ}$ (corresponding to the fraction of the Galaxy plane visible from the LHAASO site with zenith angle $\theta \lesssim 40^\circ$), in one year LHAASO can observe each source of this region for an average time of $\sim$1800 hours ($\sim$5 hours per day). 
An IACT, with its limited FoV, must scan the whole region with different pointings. The number of pointings determines the maximum observation time that can be dedicated to any source. 
Assuming a FoV of radius 5$^{\circ}$ with a decrease of sensitivity of about 50$\%$ at a distance of 3$^{\circ}$ from the center and a step for  different pointings  of $4^{\circ}$, the galactic survey requires approximately 100 pointings.
Assuming a total observation time of $\sim$1300 hours/year, a full year dedicated to the survey allows an  exposure of $\sim$13 hours for each galactic source.
The difference in sky survey capabilities between EAS-arrays and IACTs is more impressive in case of an {\it all sky survey}. Assuming a region of 2$\pi$ sr to be scanned, the number of pointings is approximately $\sim$1600, and every location would be
observed for less than one hour. 

The reduced observation time causes an increase of the minimum detectable flux, as shown in the blue band of Fig.\ref{fig:lhaaso-skysurvey}, where the integral sensitivities for sky survey of CTA-South \cite{cta-survey} and LHAASO are compared.
The lower limit of the band refers to a Galactic plane survey and the upper limit to an {\it all sky survey} of $\pi$ sr \cite{cta-survey,hiscore}.
The different points refer to upper limits set by different experiments to high energy gamma-ray emission in the Northern hemisphere. 

In conclusion, the two techniques (EAS detectors and Cherenkov telescopes) have both great scientific potential, and should be seen as complementary approaches in the exploration of different aspects, and different energy ranges, of the $\gamma$-ray emission.
On the contrary, CR studies are a peculiar characteristics of EAS-arrays.

LHAASO will perform accurate measurements of the high energy tails of emission spectra for the majority of the known TeV galactic sources visible from its location.
To give a quantitative idea of the LHAASO capabilities, it is useful to compare the detector sensitivity with the fluxes of such sources. Out of 84 sources crossing the detector FoV with a zenith angle less than 40$^{\circ}$, 36 can be reasonably considered galactic. For 35 out of these the flux has been measured and reported in \cite{tevcat} and for 24 of them a spectral index is available, ranging from 1.75 to 3.1, with an average value of 2.4 \cite{vernetto}.

\begin{figure}
  \begin{center}
 \includegraphics[width=0.8\columnwidth]{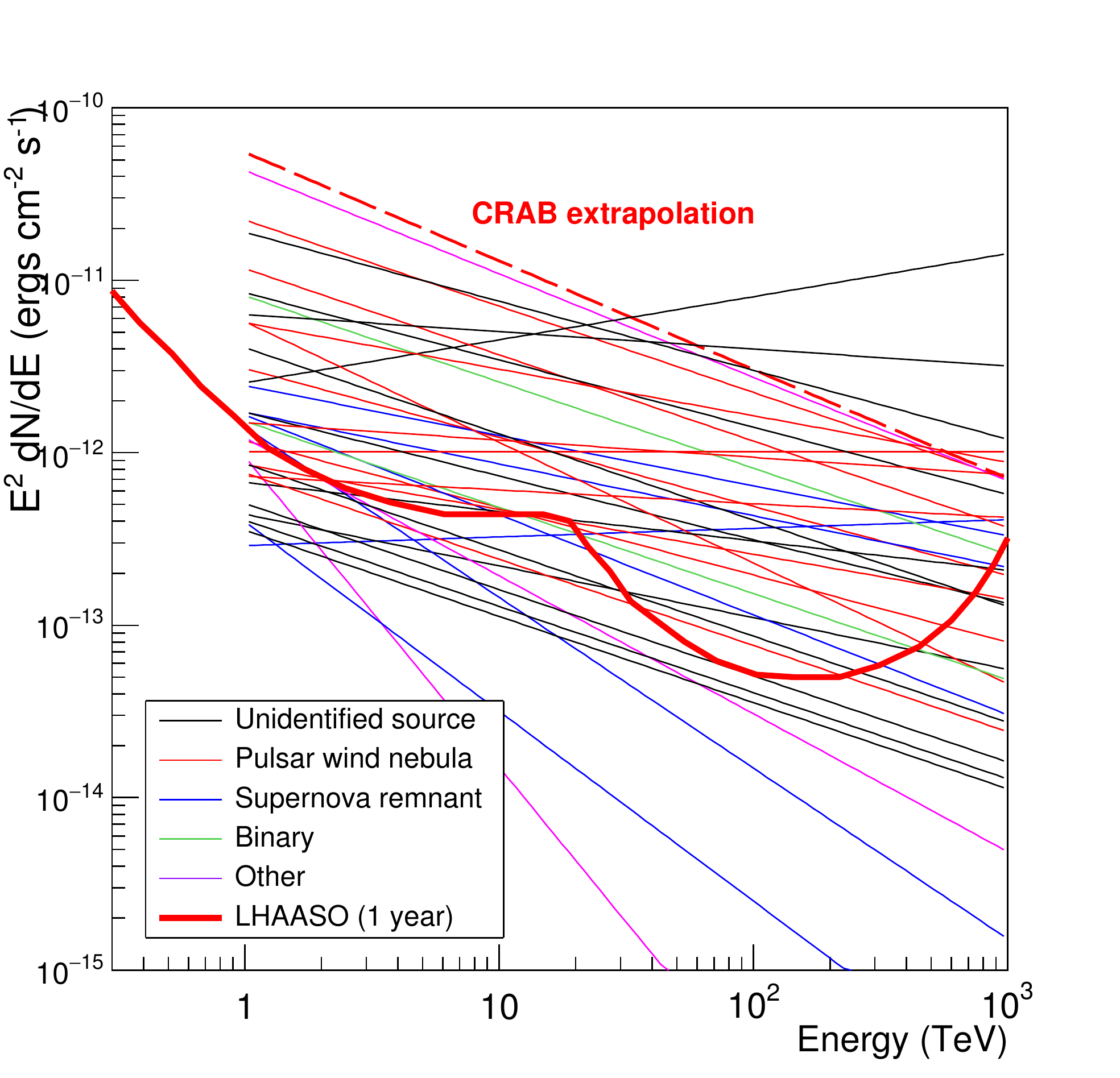}
    \caption{Differential spectra (multiplied by E$^2$) of gamma ray sources 
visible by LHAASO extrapolated to 1 PeV, compared to the LHAASO sensitivity. 
The dashed red line represents the Crab nebula flux, as measured by ARGO-YBJ
\cite{argo-crab}, extrapolated to 1 PeV.}
    \label{fig:alltev_spectra}
  \end{center}
\end{figure}

Fig. \ref{fig:alltev_spectra} shows the spectra of 35 objects extrapolated to 1 PeV (with the same spectral index measured in the TeV region) compared to the LHAASO one-year sensitivity \cite{vernetto}.
The spectral index has been set to 2.5 for the sources without an available spectral measurement.
Detailed calculations for each source are needed but the purpose of this figure is to show that the flux of almost all the considered sources is above the LHAASO sensitivity. These high energy data are  likely to play a crucial role for the understanding of the properties of the sources.

\begin{figure}
  \begin{center}
 \includegraphics[width=0.8\columnwidth]{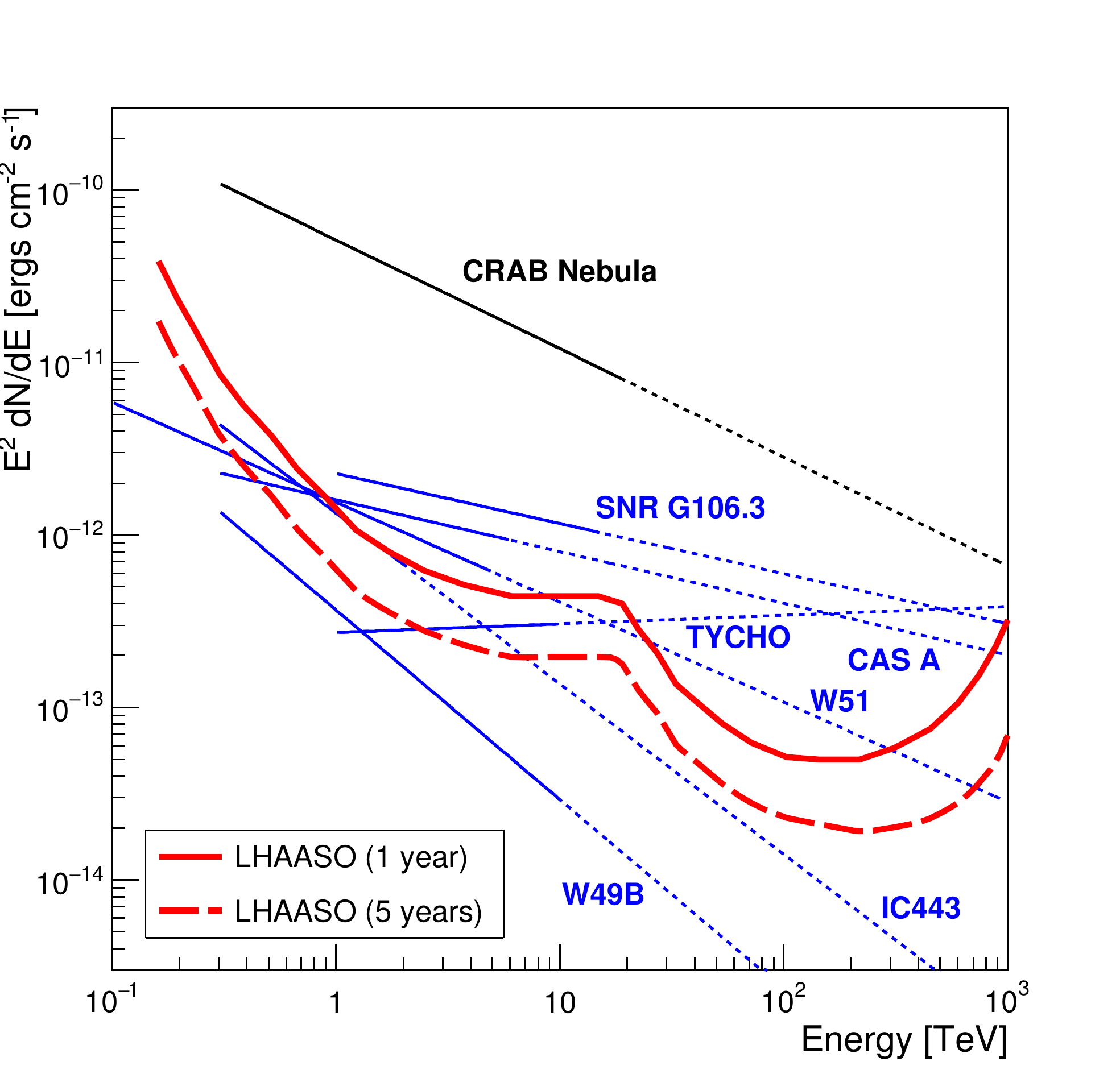}
    \caption{Differential spectra (multiplied by E$^2$) of six supernova remnants 
visible by LHAASO (solid blue lines), extrapolated to 1 PeV (dotted blue
lines), compared to the LHAASO sensitivity. 
The black line represent the Crab nebula flux, as measured by ARGO-YBJ \cite{argo-crab}, extrapolated to 1 PeV.}
    \label{fig:snr}
  \end{center}
\end{figure}

Among galactic sources, shell SNRs are the most interesting ones to be studied at high energy because they are believed to be the sources of galactic CRs.
In the LHAASO FoV there are six shell SNRs (Thyco, CAS A, W51, IC443, W49B and SNR G106.3+2.7). 
Fig.\ref{fig:snr} shows their spectra extrapolated to 1 PeV. Four of them have fluxes higher than the LHAASO 1-year sensitivity. 

Besides the observation of known sources, given the LHAASO capabilities in sky survey, new galactic sources will likely be discovered at high energy, since objects with fluxes at 1 TeV below the current instruments sensitivity but with hard spectra (i.e. spectral index $<$2) would be easily detectable by LHAASO above $\sim$10 TeV.

A wide FoV detector such as LHAASO is particularly sensitive to extended sources (molecular clouds, galaxies, dwarf galaxies and galay clusters), being an ideal detector for observations of diffuse $\gamma$-ray emission from the Galactic Plane.
Ground based gamma-ray detectors, both EAS arrays and IACTs, lose sensitivity observing extended sources. 
When the source size is large compared to PSF (or FoV) the sensitivity for point sources is reduced by a factor $\sigma_{det}/\sigma_{source}$, where $\sigma_{det}$ is the detector PSF and $\sigma_{source}$ the source diameter.
Air shower arrays, however, due to larger PSF and FoV, are less affected than Cherenkov instruments.

\begin{figure}
  \begin{center}
 \includegraphics[width=0.8\columnwidth]{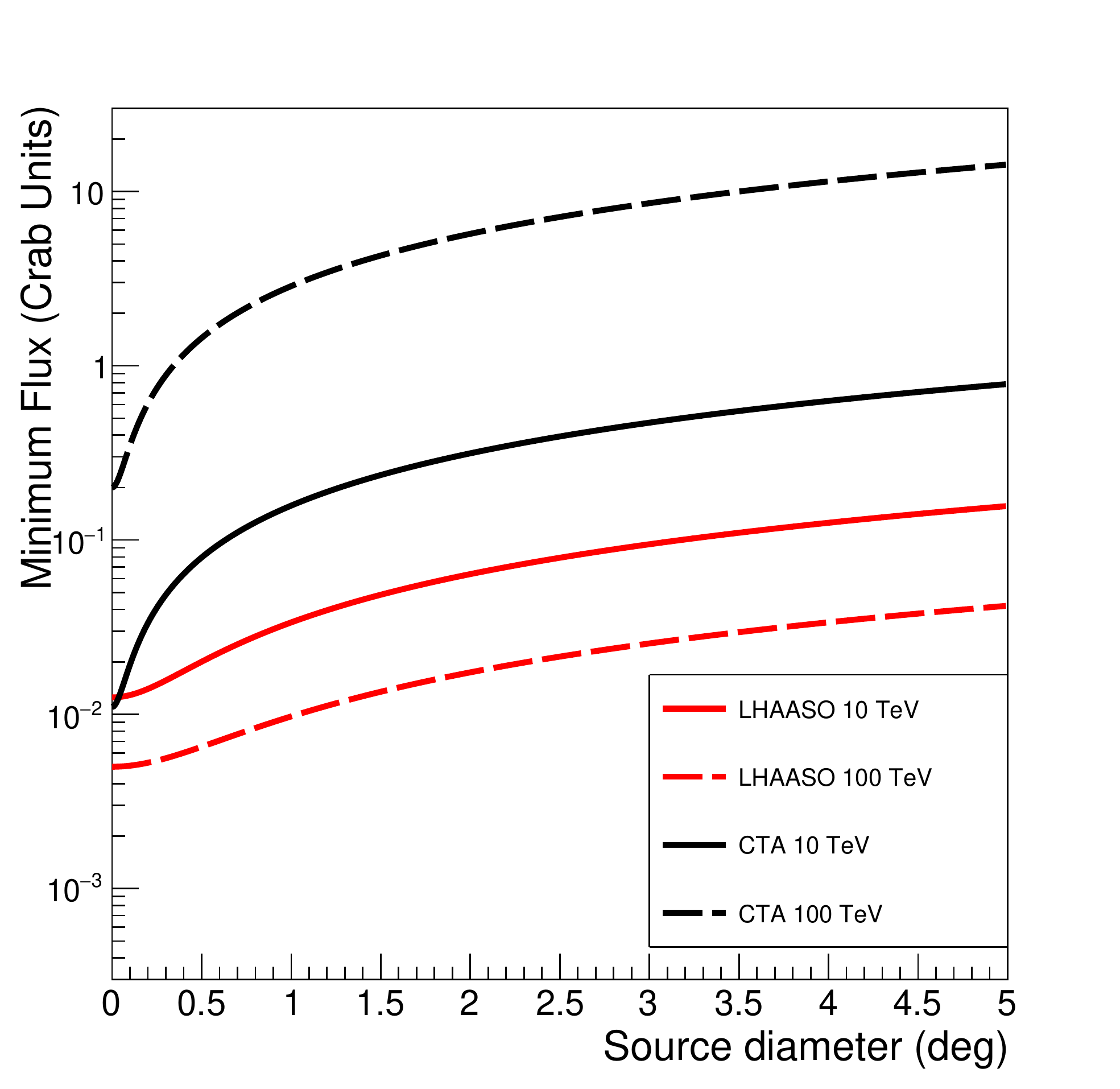}
    \caption{The minimum integral flux (in Crab units) detectable by LHAASO and CTA-South as a function of the source angular diameter, for two different photon energies.}
    \label{fig:extended}
  \end{center}
\end{figure}

As an example, in Fig. \ref{fig:extended} the LHAASO minimum detectable integral flux (in Crab units) in one year is shown as a function of the source diameter for two different photon energies. For comparison the minimum flux expected by a Cherenkov telescope like CTA-South in 50 h is shown. As it can be seen, an EAS array with the LHAASO characteristics is well suitable for high sensitivity study of extended sources, in particular for energies above tens of TeV.

The space distribution of diffuse $\gamma$-rays can trace the location of the CR sources and the distribution of interstellar gas.
In fact, this emission can be produced by protons and nuclei via the decay of $\pi^0$ produced in hadronic interactions with interstellar gas.
The study of the diffuse flux at 30 --100 TeV energies would be of extreme importance to understand the propagation and the confinement of the parent CRs in the Galaxy and their source distribution.
In addition, the observation of a knee in the energy spectrum of diffuse emission would provide a complementary way to investigate the elemental composition in the knee range. Observing a location dependence of the knee energy would provide important clues on the nature of the knee, as do similar measurement for individual sources of CRs.

A rough evaluation of the LHAASO sensitivity to the galactic diffuse flux can be obtained by multiplying the point source sensitivity given in Fig. \ref{fig:lhaaso_sens} by the correction factor $f$ = ($\Omega_{PSF} \Omega_{GP}$)$^{-1/2}$, where $\Omega_{PSF}$ is the observation angular window, related to the detector PSF and $\Omega_{GP}$ in the solid angle of the galactic plane region to be studied.
According to this simple calculation, the 5 sigma minimum flux detectable by LHAASO in one year in the longitude interval 25$^\circ$--100$^\circ$ would be F$_{min}\sim$7$\times$10$^{-16}$ photons cm$^{-2}$ s$^{-1}$ TeV$^{-1}$ sr$^{-1}$ at 100 TeV, a 
factor $\sim$3 lower than the extrapolation of the flux measured at 1 TeV by ARGO-YBJ \cite{argo-diffuse}.
This implies that LHAASO will be able to study the properties of gamma rays produced by the interaction of CRs with energy up to the all-particle knee and to measure the knee in the diffuse energy spectrum corresponding to the ARGO-YBJ observation of the proton knee at $\sim$700 TeV \cite{hybrid15}.

\section{Conclusions}

The LHAASO experiment is expected to be the most sensitive detector to study the multi-TeV $\gamma$-ray sky opening for the first time the 10$^2$--10$^3$ TeV range to the direct observations of the high energy CR sources.
This unique detector will be capable of continuously surveying the gamma-ray sky for steady and transient sources from about 100 GeV to PeV energies.
When new wavelength bands are explored in astronomy, previously unknown sources and unknown types of sources are discovered. The discovery of new classes of objects, unobserved at other wavelengths, is a major strength of all-sky monitors. 

The installation of an unprecedented muon detection area ($\sim$42,000 m$^2$), together with an array of wide field of view Cerenkov telescopes, will allow high resolution selection of showers induced by different primary masses in the range 10$^{12}$ -- 10$^{17}$ eV and a detailed study of the role of the hadronic interaction models.





\end{document}